\documentclass[preprint,jmp]{revtex4-1}
\usepackage{amssymb}
\usepackage{amsmath}
\usepackage{amsfonts}

\begin{document}

\title{From the generalized Morse potential to a unified treatment of the $D$%
-dimensional singular harmonic oscillator and singular Coulomb potentials{\footnote {To appear in Journal of Mathematical Chemistry}} \\
}
\author{Pedro H. F. Nogueira}
\email{pedrofusconogueira@gmail.com}
\author{Antonio S. de Castro}
\email{castro@pq.cnpq.br}\affiliation{Universidade Estadual Paulista - Campus de Guaratinguet\'{a},
Departamento de F\'{\i}sica e Qu\'{\i}mica, 12516-410 Guaratinguet\'{a} SP,
Brazil}

\begin{abstract}
Bound-state solutions of the singular harmonic oscillator and singular
Coulomb potentials in arbitrary dimensions are generated in a simple way
from the solutions of the one-dimensional generalized Morse potential. The
nonsingular harmonic oscillator and nonsingular Coulomb potentials in
arbitrary dimensions with their additional accidental degeneracies are
obtained as particular cases. Added bonuses from these mappings are the
straightforward determination of the critical attractive singular potential,
the proper boundary condition on the radial eigenfunction at the origin and
the inexistence of bound states in a pure inversely quadratic potential.
\end{abstract}

\maketitle

\section{Introduction}

Some exactly soluble systems with importance in atomic and molecular physics
have been approached in the literature on quantum mechanics with a myriad of
methods. Among such systems is the Morse potential $a(e^{-\alpha
x}-2e^{-2\alpha x})$ \cite{b1mor}-\cite{b1don3}, the $D$-dimensional
pseudoharmonic potential $a\left( x/b-b/x\right) ^{2}$ \cite{b1gol}-\cite%
{b1haa}, \cite{b2wei}-\cite{b2tez}, and the $D$-dimensional Kratzer-Fues
potential $a\left( b^{2}/x^{2}-2b/x\right) $ and its modified version $%
a\left( b^{2}/x^{2}-b/x\right) $ \cite{b1gol}-\cite{b1haa}, \cite{b1flu},
\cite{b2ikh2}-\cite{b3mol}. More general exactly soluble systems have also
been appreciated: the generalized Morse potential $Ae^{-\alpha
x}+Be^{-2\alpha x}$ \cite{b2tez}, \cite{b4bar}-\cite{b4ard1}, the singular
harmonic oscillator $Ax^{2}+Bx^{-2}$ \cite{b1lan}-\cite{b1haa}, \cite{b1don3}%
, \cite{b4bag}, \cite{b5con}-\cite{b5asc} and the singular Coulomb potential
\thinspace $Ax^{-1}+Bx^{-2}$ \cite{con}-\cite{b1haa}, \cite{b4bag}, \cite%
{b5con}, \cite{b5don1}, \cite{b5ikh}, \cite{b6hal}-\cite{das}.

In a recent paper \cite{jmc}, it was shown that the Schr\"{o}dinger equation
for all those exactly solvable problems mentioned above can be reduced to
the confluent hypergeometric equation in such a way that it can be solved
via Laplace transform method with closed-form eigenfunctions expressed in
terms of generalized Laguerre polynomials. Connections between the Morse and
those other potentials have also been reported. The three-dimensional
Coulomb potential has been mapped into the one-dimensional Morse potential
and into the three-dimensional singular Coulomb potential via change of
function and variable \cite{lang}. The Morse potential with particular
parameters has been mapped into the two-dimensional harmonic oscillator \cite%
{mon} and into the three-dimensional Coulomb potential \cite{lee}. Later,
the generalized Morse potential was mapped into the three-dimensional
harmonic oscillator and Coulomb potentials \cite{hay}-\cite{sun}.
Furthermore, a certain mapping between the Morse potential with particular
parameters and the three-dimensional Kratzer-Fues potential has been found
with fulcrum on the algebra $so\left( 2,1\right) $ and its representations
\cite{coo}. Although the Morse and the Kratzer-Fues potentials are
well-known systems, the map connecting them was used recently to obtain the
Wigner distribution functions for the Kratzer-Fues potential from the Wigner
distribution functions of the Morse potential \cite{sta}.

In this paper, an alternative and more general approach for the mapping is
developed. We show that bound-state solutions of the singular harmonic
oscillator and singular Coulomb potentials in arbitrary dimensions can be
generated in a simple way from the bound states of the one-dimensional
generalized Morse potential via Langer transformation \cite{lang}. Links
with the nonsingular harmonic oscillator and nonsingular Coulomb potentials
in arbitrary dimensions with their additional accidental degeneracies are
obtained as particular cases. Added bonuses from these interrelationships
are the straightforward determination of the critical attractive singular
potential (that one which avoids the famous \textquotedblleft fall of a
particle to the centre\textquotedblright\ \cite{b1lan}) and the proper
boundary condition on the radial eigenfunction at the origin (that one which
excludes spurious solutions coming from the Laplacian operator (see, e.g.
\cite{spec}-\cite{khe})). As a mere epiphenomenon of our approach, it is
shown that a pure inversely quadratic potential can not hold bound states.

In Sec. II we present a detailed analysis of the bound-state solutions in a
one-dimensional generalized Morse potential. In Sec. III we present a few
relevant properties of the Schr\"{o}dinger equation in $D$ dimensions for
spherically symmetric potentials. We then proceed to show that the bound
states in the singular harmonic oscillator and singular Coulomb potentials
are linked to the bound states in the generalized Morse potential. Final
remarks comprise Sec. IV.

\section{Bound states in a generalized Morse potential}

The time-independent Schr\"{o}dinger equation is an eigenvalue equation for
the characteristic pair $(E,\psi )$ with $E\in
\mathbb{R}
$. For a particle of mass $m$ embedded in a one-dimensional potential $%
V\left( x\right) $ it is given by
\begin{equation}
\frac{d^{2}\psi \left( x\right) }{dx^{2}}+\frac{2m}{\hbar ^{2}}\left[
E-V\left( x\right) \right] \psi \left( x\right) =0,  \label{sch}
\end{equation}%
where $\hbar $ is Planck's constant, and $\int_{-\infty }^{+\infty
}dx\,|\psi |^{2}=1$ for bound states. For the generalized Morse potential%
\begin{equation}
V(x)=V_{1}e^{-\alpha x}+V_{2}e^{-2\alpha x},\quad \alpha >0,  \label{morse}
\end{equation}%
the substitution%
\begin{equation}
\xi =\frac{2\sqrt{2mV_{2}}\,e^{-\alpha x}}{\hbar \alpha }  \label{xi}
\end{equation}%
and the definitions%
\begin{equation}
s=\frac{\sqrt{-2mE}}{\hbar \alpha },\quad a=\frac{mV_{1}}{\hbar \alpha \sqrt{%
2mV_{2}}}+s+\frac{1}{2}  \label{k}
\end{equation}%
convert Eq. (\ref{sch}) into%
\begin{equation}
\frac{d^{2}\psi \left( \xi \right) }{d\xi ^{2}}+\frac{1}{\xi }\frac{d\psi
\left( \xi \right) }{d\xi }+\left( -\frac{1}{4}+\frac{s-a+1/2}{\xi }-\frac{%
s^{2}}{\xi ^{2}}\right) \psi \left( \xi \right) =0,  \label{whi}
\end{equation}%
whose solutions have asymptotic limits expressed as $\psi \left( \xi \right)
\underset{|\xi |\rightarrow 0}{\rightarrow }\xi ^{\pm s}$ and $\psi \left(
\xi \right) \underset{|\xi |\rightarrow \infty }{\rightarrow }e^{\pm \xi /2}$%
. On account of the normalization condition, $\int_{0}^{\infty }d|\xi
|\,|\psi \left( \xi \right) |^{2}/|\xi |=\alpha $, one has that $\psi $
behaves like $\xi ^{s}$ as $|\xi |\rightarrow 0$ and like $e^{-\xi /2}$ as $%
|\xi |\rightarrow \infty $ with $\xi \in
\mathbb{R}
$ ($V_{2}>0$) and $s>0$ ($E<0$). We write $\psi =e^{-\xi /2}\xi ^{s}w$,
where $w$ satisfies Kummer's equation
\begin{equation}
\xi \frac{d^{\,2}w\left( \xi \right) }{d\xi ^{2}}+\left( 2s+1-\xi \right) \,%
\frac{dw\left( \xi \right) }{d\xi }-aw\left( \xi \right) =0  \label{kum}
\end{equation}%
with general solution expressed as
\begin{equation}
w\left( \xi \right) =AM\left( a,2s+1,\xi \right) +BU\left( a,2s+1,\xi
\right) .  \label{solw}
\end{equation}%
Here, $A$ and $B$ are arbitrary constants, $M\left( a,b,z\right)
=\,_{1\!}F_{1}\left( a,b,z\right) $ is the confluent hypergeometric function
\cite{leb}, and
\begin{equation}
U(a,b,z)=\frac{\pi }{\sin \pi b}\left[ \frac{M\left( a,b,z\right) }{\Gamma
\left( 1+a-b\right) }-z^{1-b}\frac{M\left( 1+a-b,2-b,z\right) }{\Gamma
\left( a\right) \Gamma \left( 2-b\right) }\right] ,
\end{equation}%
where $\Gamma \left( z\right) $ is the gamma function. One has to search
particular solutions of Eq. (\ref{kum}) such that $w\left( \xi \right)
\underset{\xi \rightarrow 0}{\rightarrow }C$ and $w\left( \xi \right)
\underset{\xi \rightarrow \infty }{\rightarrow }\xi ^{\alpha _{1}}e^{\alpha
_{2}\xi ^{\alpha _{3}}}$, where $C$ is a nonvanishing constant, $\alpha _{1}$
and $\alpha _{2}$ are arbitrary constants, and $\alpha _{3}<1$. This occurs
because $\xi ^{\alpha _{1}}e^{\alpha _{2}\xi ^{\alpha _{3}}-\xi
/2}\rightarrow e^{-\xi /2}$ as $\xi \rightarrow \infty $. For the reason
that \cite{leb}
\begin{equation}
M\left( a,b,z\right) \underset{|z|\rightarrow 0}{\rightarrow }1,\quad
U\left( a,b,z\right) \underset{|z|\rightarrow 0}{\rightarrow }\frac{\Gamma
\left( b-1\right) z^{1-b}}{\Gamma \left( a\right) }\quad \text{\textrm{for}}%
\quad b\neq 1,
\end{equation}%
one has $B=0$. On the other hand \cite{leb},%
\begin{equation}
\frac{M\left( a,b,z\right) }{\Gamma \left( b\right) }\underset{%
|z|\rightarrow \infty }{\rightarrow }\frac{e^{i\pi a}z^{-a}}{\Gamma \left(
b-a\right) }+\frac{e^{z}z^{a-b}}{\Gamma \left( a\right) },\quad -\pi /2<\arg
z<3\pi /2,  \label{333}
\end{equation}%
so that $w$ diverges as $e^{\xi }$ for large $\xi $. Due to the poles of the
gamma function in (\ref{333}), this bad behaviour can remedied making $%
-a=n\,\in
\mathbb{N}
$. It follows from (\ref{k}) that $V_{1}<0$ and therefore the generalized
Morse potential is able to hold bound states only if it has a well structure
($V_{1}<0$ and $V_{2}>0$). Furthermore, $M\left( -n,b,z\right) $ is
proportional to the generalized Laguerre polynomial $L_{n}^{\left(
b-1\right) }\left( z\right) $, a polynomial of degree $n$ \cite{leb}.
Therefore,%
\begin{equation}
\psi _{n}\left( \xi \right) =N_{n}\,\xi ^{s}e^{-\xi /2}L_{n}^{\left(
2s\right) }\left( \xi \right) ,
\end{equation}%
where $N_{n}$ is a normalization constant. Substituting $a=-n$ in Eq. (\ref%
{k}), one finds the quantization condition%
\begin{equation}
n+s+\frac{1}{2}=\frac{m|V_{1}|}{\hbar \alpha \sqrt{2mV_{2}}},
\end{equation}%
and because $s>0$ one gets%
\begin{equation}
n<\frac{m|V_{1}|}{\hbar \alpha \sqrt{2mV_{2}}}-\frac{1}{2}.  \label{cond}
\end{equation}%
This restriction on $n$ limits the number of allowed states and requires $%
m|V_{1}|/\left( \hbar \alpha \sqrt{2mV_{2}}\right) >1/2$ to make the
existence of a bound state possible. Finally, the solution of the
quantization condition is expressed as%
\begin{equation}
E_{n}=-\frac{V_{1}^{2}}{4V_{2}}\left[ 1-\frac{\hbar \alpha \sqrt{2mV_{2}}}{%
m|V_{1}|}\left( n+\frac{1}{2}\right) \right] ^{2}.  \label{en}
\end{equation}%
These results for the generalized Morse potential is in agreement with those
ones obtained in Ref. \cite{jmc} via Laplace transform method.

\section{Bound states in D dimensions}

The $D$-dimensional time-independent Schr\"{o}dinger equation is expressed
as (see, e.g. \cite{b1don3}, \cite{cha})

\begin{equation}
-\frac{\hbar ^{2}}{2m}\nabla _{D}^{2}\psi \left( \vec{r}\right) +V\left(
\vec{r}\right) \psi \left( \vec{r}\right) =\varepsilon \psi \left( \vec{r}%
\right) ,  \label{eq1}
\end{equation}%
where $\nabla _{D}^{2}$ is the $D$-dimensional Laplacian operator. In
spherical coordinates $\vec{r}=\left( r,\Omega \right) $. Here, $r=|\vec{r}%
|\,\in \lbrack 0,\infty )$, and $\Omega $ denotes a set of $D-1$ angular
variables. Eq. (\ref{eq1}) is an eigenvalue equation for the characteristic
pair ($\varepsilon ,\psi $) with $\varepsilon \in
\mathbb{R}
$ and $\int d\tau \,|\psi |^{2}=1$ for bound states. In this last formula $%
d\tau =r^{D-1}drd\Omega $ is the volume element and the integral is taken
over the whole hyperspace.

For spherically symmetric potentials one can write (see, e.g. \cite{b1don3},
\cite{cha})%
\begin{equation}
\psi \left( \vec{r}\right) =r^{\left( 1-D\right) /2}\,u(r)Y\left( \Omega
\right) ,  \label{F}
\end{equation}%
where $u$ obeys the radial equation%
\begin{equation}
\frac{d^{2}u\left( r\right) }{dr^{2}}+\frac{2m}{\hbar ^{2}}\left[
\varepsilon -V\left( r\right) -\frac{L\left( L+1\right) \hbar ^{2}}{2mr^{2}}%
\right] u\left( r\right) =0  \label{Eq}
\end{equation}%
with $\int_{0}^{\infty }dr\,|u|^{2}=1$ for bound-state solutions, and%
\begin{equation}
L=l+\left( D-3\right) /2\quad \text{or}\quad L=-l-\left( D-1\right) /2,
\label{EQ2}
\end{equation}%
in which $l=0,1,2,\ldots $ In (\ref{F}), $Y$ denotes the normalized
hyperspherical harmonics ($\int d\Omega \,|Y|^{2}=1$) labeled by $D-1$
quantum numbers:%
\begin{eqnarray}
l_{D-1} &=&l,\quad l_{D-2}=0,1,2,\ldots ,l_{D-1},\quad l_{D-3}=0,1,2,\ldots
,l_{D-2},  \notag \\
&&\vdots  \notag \\
l_{4} &=&0,1,2,\ldots ,l_{5},\quad l_{3}=0,1,2,\ldots ,l_{4},\quad
l_{2}=0,1,2,\ldots ,l_{3}, \\
&&  \notag \\
l_{1} &=&-l_{2},-l_{2}+1,\ldots ,+l_{2}-1,+l_{2}.  \notag
\end{eqnarray}%
Hence, the essential degeneracy of the spectrum for a given $l$ is expressed
by \cite{cha}%
\begin{equation}
d_{l}\left( D\right) =\frac{\left( D+2l-2\right) \left( D+l-3\right) !}{%
l!\left( D-2\right) !}.  \label{Q3}
\end{equation}

With potentials expressed as

\begin{equation}
V(r)=Zr^{\delta }+\frac{\hbar ^{2}\beta }{2mr^{2}},  \label{pot}
\end{equation}%
the Langer transformation \cite{lang}%
\begin{equation}
u=\sqrt{r/r_{0}}\,\phi ,\quad r/r_{0}=e^{-\Lambda \alpha x},
\end{equation}%
with $r_{0}>0$ and $\Lambda >0$, transmutes the radial equation (\ref{Eq})
into%
\begin{equation}
\frac{d^{2}\phi \left( x\right) }{dx^{2}}+\frac{2m}{\hbar ^{2}}\left\{ -%
\frac{\left( \hbar \Lambda \alpha S\right) ^{2}}{2m}-\left( \Lambda \alpha
r_{0}\right) ^{2}\left[ Zr_{0}^{\delta }e^{-\Lambda \alpha \left( \delta
+2\right) x}-\varepsilon e^{-2\Lambda \alpha x}\right] \right\} \phi \left(
x\right) =0,  \label{sch2}
\end{equation}%
with%
\begin{equation}
S=\sqrt{\beta +\left( L+1/2\right) ^{2}}.  \label{etil}
\end{equation}%
At this point, it is instructive to note that $S$ is insensible to the
different choices of $L$ as prescribed by (\ref{EQ2}). Besides that, Eq. (%
\ref{sch2}) is precisely the Schr\"{o}dinger equation for the `Morse
potential' with $V_{1}=0$ or $V_{2}=0$ when (\ref{pot}) is the pure
inversely quadratic potential ($\delta =0$ or $\delta =-2$). In this case
there is no bound-state solution. Nevertheless, a connection with the bound
states of the generalized Morse potential, with $\int_{-\infty }^{+\infty
}dx\,e^{-2\Lambda \alpha x}|\phi |^{2}=\left( \Lambda \alpha r_{0}\right)
^{-1}$, might be reached if the pair $(\delta ,\Lambda )$ is equal to $%
(2,1/2)$ or $(-1,1)$. As an immediate consequence of the mapping for bound
states $S^{2}>0$. Thus,
\begin{equation}
\beta >-\left( D-2\right) ^{2}/4.  \label{beta}
\end{equation}%
Furthermore, the existence of bound states also demands $\phi \left(
x\right) \underset{x\rightarrow +\infty }{\rightarrow }e^{-\Lambda \alpha
S\,x}$ with $\Lambda S>0$, in such a way that%
\begin{equation}
u\left( r\right) \underset{r\rightarrow 0}{\rightarrow }r^{1/2+S}.
\label{ur}
\end{equation}%
The above restriction on the coupling constant $\beta $ and the boundary
condition $u\left( 0\right) =0$ represent important pieces for the
determination of bound states. The first one excludes strongly attractive
singular potentials and can be obtained by recurring to a regularization of
the potential at the origin (see, e.g. \cite{b1lan}). The second one,
well-grounded even for nonsingular potentials, can be legitimated by ruling
out the Dirac delta function $\delta \left( \vec{r}\right) $ coming from the
Laplacian operator in (\ref{eq1}) (see, e.g. \cite{spec}-\cite{khe}).

\subsection{The singular harmonic oscillator}

With $\delta =2$ plus the definition $Z=m\omega ^{2}/2$, the potential (\ref%
{pot}) is written as%
\begin{equation}
V(r)=\frac{1}{2}m\omega ^{2}r^{2}+\frac{\hbar ^{2}\beta }{2mr^{2}}.
\end{equation}%
In order to complete the identification of the bound-state solutions with
those ones from the generalized Morse potential one must choose $\Lambda
=1/2 $, $V_{1}=-\alpha ^{2}r_{0}^{2}\varepsilon /4$ and $V_{2}=\alpha
^{2}r_{0}^{4}m\omega ^{2}/8$. For the reason that $V_{1}<0$ and $V_{2}>0$
one can see that bound-state solutions require $\varepsilon >0$ and $\omega
^{2}>0$, respectively, and choosing $\omega >0$ one can write%
\begin{equation}
\xi =m\omega r^{2}/\hbar .
\end{equation}%
Furthermore, (\ref{cond}) implies $\varepsilon >2\hbar \omega \left(
n+1/2\right) $. Using (\ref{en}) and (\ref{etil}) one can write the complete
solution of the problem as%
\begin{eqnarray}
\varepsilon _{nL} &=&\hbar \omega \left( 2n+1+S\right) ,  \notag \\
&& \\
u_{nL}(r) &=&A_{nL}r^{1/2+S}e^{-m\omega r^{2}/\left( 2\hbar \right)
}L_{n}^{\left( S\right) }\left( m\omega r^{2}/\hbar \right) .  \notag
\end{eqnarray}%
When $\beta =0$, the case of a pure harmonic oscillator, one can write%
\begin{equation}
\varepsilon _{N}=\hbar \omega \left( N+D/2\right) ,\quad N=0,1,2,\ldots ,
\end{equation}%
where $N=2n+l$. The radial eigenfunction $u$, though, is labelled with the
quantum numbers $N$ and $l$, with $l$ even (odd) for $N$ even (odd) and $%
l\leq N$.

\subsection{The singular Coulomb potential}

Now, $\delta =-1$ and%
\begin{equation}
V(r)=\frac{Z}{r}+\frac{\hbar ^{2}\beta }{2mr^{2}}.
\end{equation}%
Comparison of the bound states with those ones from the generalized Morse
potential is done by choosing $\Lambda =1$, $V_{1}=\alpha ^{2}r_{0}Z$ and $%
V_{2}=-\alpha ^{2}r_{0}^{2}\varepsilon $. The conditions $V_{1}<0$ and $%
V_{2}>0$ imply $Z<0$ and $\varepsilon <0$, respectively. Now,
\begin{equation}
\xi =2\sqrt{2m|\varepsilon |}\,r/\hbar
\end{equation}%
and (\ref{cond}) implies $\varepsilon >-\hbar ^{2}/[2ma^{2}\left(
n+1/2\right) ^{2}]$. Here, $a=\hbar ^{2}/\left( m|Z|\right) $. Using (\ref%
{en}) and (\ref{etil}) one can write%
\begin{eqnarray}
\varepsilon _{nL} &=&-\frac{\hbar ^{2}}{2ma^{2}\left( n+1/2+S\right) ^{2}},
\notag \\
&& \\
u_{nL}\left( r\right) &=&B_{nL}r^{1/2+S}e^{-r/\left[ a\left( n+1/2+S\right) %
\right] }L_{n}^{\left( 2S\right) }\left( 2r/\left[ a\left( n+1/2+S\right) %
\right] \right) .  \notag
\end{eqnarray}%
In the case of a pure Coulomb potential ($\beta =0$), one can write%
\begin{equation}
\varepsilon _{N}=-\frac{\hbar ^{2}}{2ma^{2}\left[ N+\left( D-3\right) /2%
\right] ^{2}},\quad N=1,2,3,\ldots
\end{equation}%
Here $N=n+l+1$ and the radial eigenfunction $u$ is labelled with the quantum
numbers $N$ and $l$, with $l$ $\leq N-1$.

\section{Final remarks}

We have shown that the complete infinite sets of bound-state solutions of
the singular harmonic oscillator and singular Coulomb potentials (and their
higher degenerate nonsingular counterparts) in arbitrary dimensions can be
extracted from the finite set of bound-state solutions of the
one-dimensional generalized Morse potential in a simple way. Surprisingly,
the determination of the critical coupling constant $\beta _{c}=-\left(
D-2\right) ^{2}/4$ as well as the proper boundary condition $u\left(
0\right) =0$ emerged in a natural manner. As a by-product, we have shown
that there is no bound state in a pure inversely quadratic potential.

\begin{acknowledgments}
We thank Prof. Dr. M. G. Garcia for useful discussions. This work was supported in part by means of funds provided by  FAPESP and CNPq (grant 304743/2015-1).
\end{acknowledgments}

\end{document}